\documentclass[
    aps,
    prapplied,    
    reprint,
    longbibliography,
    amsmath,
    amssymb,
    superscriptaddress,
]{revtex4-2}

\usepackage{graphicx}
\usepackage{dcolumn}
\usepackage{bm}
\usepackage{bbm}
\usepackage[caption=false]{subfig}
\usepackage[colorlinks=true, citecolor=blue, urlcolor=blue, linkcolor=blue]{hyperref}
\usepackage{url}
\usepackage{float}
\usepackage{textcomp}
\usepackage{color,soul}
\soulregister{\cite}{7}

\begin{document}


\title{Dispersive Hong-Ou-Mandel Interference with Finite Coincidence Windows}


\author{T.J. Walstra}
\affiliation{Department of Applied Nanophotonics, University of Twente, 7500 AE Enschede, The Netherlands}

\author{A.J. Hasenack}
\affiliation{Department of Mathematics and Computer Science, Eindhoven University of Technology, 5600 MB Eindhoven, The Netherlands}

\author{D.J. de Ruiter}
\affiliation{Department of Applied Nanophotonics, University of Twente, 7500 AE Enschede, The Netherlands}

\author{P.W.H. Pinkse}
\email{p.w.h.pinkse@utwente.nl}
\affiliation{Department of Applied Nanophotonics, University of Twente, 7500 AE Enschede, The Netherlands}

\author{T.D. Bradley}
\affiliation{High-Capacity Optical Transmission Laboratory, Eindhoven University of Technology, 5600 MB Eindhoven, The Netherlands}

\author{B. \v{S}kori\'{c}}
\affiliation{Department of Mathematics and Computer Science, Eindhoven University of Technology, 5600 MB Eindhoven, The Netherlands}

\date{\today}


\begin{abstract}
Hong-Ou-Mandel (HOM) interference is a fundamental tool for assessing photon indistinguishability in quantum information processing. While the effect of chromatic dispersion on HOM interference has been widely studied, the interplay between dispersion and the finite detection window of realistic measurement devices remains under-explored. In this work, we demonstrate that the rectangular coincidence window inherent to modern time-tagging modules, which effectively acts as a temporal filter, breaks the standard dispersion cancellation condition and restores sensitivity to symmetric group velocity dispersion. We derive an analytical model for type-II SPDC processes that predicts a modification of the HOM dip shape, specifically the emergence of characteristic oscillations and dip broadening. We experimentally validate this theoretical framework using a ppKTP source and transmission through optical fibers of lengths up to 29\,km. The experimental data show excellent agreement with the model, confirming the presence of window-induced oscillations and allowing for the precise extraction of the fiber dispersion parameter. These findings underscore the importance of accounting for finite timing resolution in the design and characterization of dispersive quantum communication links.
\end{abstract}

\maketitle


\section{Introduction}
The Hong-Ou-Mandel (HOM) effect is a fundamental example of two-photon quantum interference and serves as an essential tool for assessing photon indistinguishability.  When two indistinguishable photons enter a balanced beam splitter at different input ports, interference between Feynman paths suppresses coincident detections at the outputs. The resulting HOM dip in the coincidence rate as a function of relative arrival time reveals information about the temporal and spectral correlations between the photons.
The effect plays a central role in quantum information processing, quantum metrology, and photonic quantum communication systems \cite{Sangouard2011, Kok2007, Nasr2003, Lyons2018, Bouchard2020}. There has also been increasing interest in HOM interference in long-distance quantum cryptography applications, such as source monitoring in quantum key distribution \cite{Duplinskiy2021, Sun2023}.

In realistic fiber-based or free-space systems, however, the temporal and spectral characteristics of photons are often modified by dispersive effects. Group velocity dispersion (GVD) and polarization mode dispersion (PMD), for instance, introduce frequency-dependent delays that broaden the temporal wave packets and reduce the degree of temporal overlap at the beam splitter. Consequently, the shape and visibility of the HOM dip become sensitive to both the magnitude of dispersion and the photon spectral correlations. Although the influence of dispersion on HOM interference has been studied before \cite{Fan2021, Okano2013, Im2021, Ryu2017, Mazzotta2016, Steinberg1992}, often within the context of dispersion cancellation, a less explored but practically significant factor is the effect of the coincidence window used in coincidence count detection. Naively, one could argue that taking the coincidence window much larger than the photon temporal widths should minimize window-induced effects. However, it has been shown that a smaller window can actually greatly increase visibility by suppressing background noise in continuous-wave setups \cite{Tsujimoto2017}. Furthermore, while \cite{Xia2023} investigates dispersive broadening with temporal filtering, it assumes a Gaussian filter profile rather than a rectangular one. Moreover, we directly match the experimental results to the dispersion relations of the SPDC crystals.

Modern coincidence measurements are typically performed using time-tagging modules with programmable coincidence windows. When photon wave packets are stretched in time by dispersion, the choice of coincidence window effectively acts as a temporal filter that selects only part of the joint temporal amplitude (JTA) of the photon pairs. As a result, the measured HOM curve no longer directly represents the intrinsic two-photon interference visibility but instead reflects a convolution of the interference signal with the detector timing response. In particular, when both photons experience similar dispersion, the broadened temporal profiles lead to an effective truncation of the interference region, producing oscillatory behavior in the observed HOM dip.

In this work, we present a detailed theoretical and experimental study of how the coincidence window of a time-tagger influences the observed HOM interference when both photons undergo dispersion. We show that the coincidence window acts as a temporal aperture that modifies the interference pattern, altering the apparent width of the dip. 

We emphasize the importance of properly accounting for coincidence window effects in the analysis of HOM measurements involving dispersive propagation, which can have implications for the design and characterization of HOM-based communication experiments.


\section{Theory}
\subsection{Time-resolved coincidence rate}
Let a two-photon state be given by $|\psi\rangle = \int dt_\text{i} \ dt_\text{s} \ \beta(t_\text{i},t_\text{s}) |t_\text{i}\rangle \otimes |t_\text{s}\rangle$ in the time domain, with s and i the signal and idler photons respectively and where $\beta$ is some (not necessarily symmetric) JTA. We delay one arm by $\tau$ yielding $\int dt_\text{i} \ dt_\text{s} \ \beta(t_\text{i},t_\text{s} - \tau) |t_\text{i}\rangle \otimes |t_\text{s}\rangle$. After passing through a beam splitter with reflection and transmission amplitudes, $\sqrt{\eta}$ and $\sqrt{1-\eta}$, and projecting on coincidences, the resulting (unnormalized) two-photon wavefunction reduces to $\psi_\tau(t_\text{i},t_\text{s}) = \eta \beta(t_\text{i},t_\text{s} - \tau) - (1-\eta)\beta(t_\text{s},t_\text{i} - \tau)$. To make the coincidence count time-resolving, we require that the second detector clicks $\sigma$ after the first, which is imposed by a $\delta(t_\text{s} - t_\text{i} - \sigma)$. In practice, two-photon JSAs are often separable in terms of sum/difference coordinates. Thus, upon rewriting $\beta(t_\text{i}, t_\text{s})$ in terms of $\tilde{\beta}(t_+, t_-) = \beta((t_+ + t_-)/\sqrt{2}, (t_+ - t_-)/\sqrt{2})$, we find $\tilde{\beta}(t_+, t_-) = \tilde{\beta}_+(t_+) \tilde{\beta}_-(t_-)$, where $\tilde{\beta}_\pm$ are normalized. Then, it can be found that the coincidence rate that is differential in $\sigma$ is given by
\begin{equation}
    \label{eq:coincidence_rate}
    \begin{split}
        c(\tau, \sigma) &= \int dt \, dt_\text{s} \, \psi^\ast_\tau(t, t_\text{s}) \psi_\tau(t, t_\text{s}) \delta(t_\text{s} - t_\text{i} - \sigma) \\ 
        &= \frac{1}{\sqrt{2}} \bigg| \eta \tilde{\beta}_- \left(\frac{\tau + \sigma}{\sqrt{2}}\right) \\
        &\quad - (1-\eta)\tilde{\beta}_- \left(\frac{\tau - \sigma}{\sqrt{2}} \right) \bigg|^2 \int d t_+ \, |\tilde{\beta}_+(t_+)|^2.
    \end{split}
\end{equation}
For a symmetric coincidence window $[-T,T]$, this leads to the coincidence rate $c(\tau) = \int_{-T}^T d\sigma \ c(\tau, \sigma)$.

\subsection{Calculation for SPDC states}
Spontaneous parametric down-conversion (SPDC) is the process where a single photon of the pump beam (p), of frequency $\omega_\text{p}$, is converted into two photons of lower energy. We consider type-II degenerate SPDC, in which we have two photons that are distinguishable only by their polarization. For the SPDC set-up described in section \ref{pa:exp_setup} this gives us the state $|\psi^{\omega_\text{p}} \rangle \propto \int d \omega_\text{i} \ d \omega_\text{s}    \ J(\omega_\text{i}, \omega_\text{s})    \ |\omega_\text{i} \rangle \otimes |\omega_\text{s} \rangle$ where $J(\omega_\text{i}, \omega_\text{s})$ is the joint spectral amplitude and can be written here as
\begin{equation*}
\begin{split}
    J(\omega_\text{i}, \omega_\text{s}) \propto \ &\text{sinc}\bigg( \frac{d}{2} \Big[ k_\text{p} (\omega_\text{i} + \omega_\text{s}) - k_\text{i} (\omega_\text{i}) - k_\text{s}(\omega_\text{s}) - \frac{2\pi}{\Lambda} \Big] \bigg) \\ 
    &\times \exp\left(-\frac{( \omega_\text{i} + \omega_\text{s} - \omega_\text{p})^2}{\sigma_\text{p}^2} \right),
\end{split}
\end{equation*}
with $\Lambda$ the poling period, $d$ the crystal length and $\sigma_\text{p}$ the pump-beam spectral uncertainty. The phase mismatch between the pump beam and the signal and idler beams is proportional to the difference of the dispersion relations. The poling period is chosen in such a way that at $\omega_\text{i} = \omega_\text{s} = \frac{1}{2} \omega_\text{p}$ the mismatch is zero. Hence, to first order around $\frac{1}{2}\omega_\text{p}$, the phase-matching term can be written as
\begin{equation*}
\begin{split}
    k_\text{p} (\omega_\text{i} &+ \omega_\text{s}) - k_\text{i} (\omega_\text{i}) - k_\text{s}(\omega_\text{s}) - \frac{2\pi}{\Lambda} \\
    &\approx \left( \omega_\text{i} - \frac{\omega_\text{p}}{2}\right) \gamma_\text{i} + \left( \omega_\text{s} - \frac{\omega_\text{p}}{2}\right) \gamma_\text{s},
\end{split}
\end{equation*}
where $\gamma_a = k_\text{p}'(\omega_\text{p}) - k_a'\left(\frac{1}{2} \omega_\text{p}\right)$ is related to the difference in the group index between the pump and idler/signal beam inside the crystal, $\Delta n_\text{g, a} := n_\text{g,0} - n_\text{g, a} = c \gamma_a$. Next, using the product expansion of the $\text{sinc}$ function up to second order around zero, we approximate $\text{sinc}(x) \approx e^{0} \cdot e^{0 \cdot x} \cdot e^{-\frac{1}{3} \cdot \frac{1}{2}x^2} = e^{-\frac{1}{6}x^2}$ and hence
\begin{equation}
    \label{eq:spdc_state}
    \begin{split}
    J(\omega_\text{i}, \omega_\text{s}) \propto \exp \bigg[ &-\frac{\left(\sum_a \left(\omega_\text{i} - \frac{1}{2}\omega_\text{p} \right) \tilde{\gamma}_a \right)^2}{\sigma_\text{pm}^2} \\
    &- \frac{( \omega_\text{i} + \omega_\text{s} - \omega_\text{p})^2}{\sigma_\text{p}^2} \bigg],
    \end{split}
\end{equation}
where we have defined $\sigma_\text{pm} =  4\sqrt{6}/((\gamma_\text{i} + \gamma_\text{s}) d)$ and $\tilde{\gamma}_\text{i} = 2 \gamma_\text{i}/(\gamma_\text{i} + \gamma_\text{s})$. If the engineered system moreover satisfies the extended phase-matching (EPM) condition \cite{Giovannetti2002}, $2 k'_\text{p}(\omega_\text{p}) \approx k'_\text{s}(\omega_\text{p}/2) + k'_\text{i}(\omega_\text{p}/2)$, then $\gamma_\text{i} \approx - \gamma_\text{s}$ and the two-photon JSA reduces to an elliptic form given by
\begin{equation*}
    \begin{split}
        J(\omega_\text{i}, \omega_\text{s}) \propto \exp \bigg[ 
            - \frac{(\omega_\text{i} - \omega_\text{s})^2}{4 r}
            &-\frac{(\omega_\text{i} + \omega_\text{s} - \omega_\text{p})^2}{4 r_\text{p}} 
        \bigg],
    \end{split}
\end{equation*}
where $r = 24/((\gamma_\text{i} - \gamma_\text{s}) d )^2$ and $r_\text{p} = \sigma_\text{p}^2 /4$. In our analysis, two relevant limits of this JSA are when the pump beam uncertainty becomes spectrally narrow, $r_\text{p} \rightarrow 0$, which corresponds to perfect frequency anti-correlation (the maximally-squeezed limit) and secondly $r_\text{p} \rightarrow r$ in which the JSA is fully separable  in $\omega_\text{i}$ and $\omega_\text{s}$. This last scenario signifies the case of no entanglement (the no-squeezing limit). 

A bandpass filter was implemented to validate the Gaussian approximation of the sinc function. To a first-order approximation, this filter can be modeled as $F(\omega) \propto \exp( - (\omega - \omega_f)^2 / 2 s )$. This Gaussian approximation is valid as long as the primary lobe of the sinc function closely matches the width of the filter, $F(\omega)$, and their central frequencies align, $\omega_f \approx \omega_\text{p}/2$.
When the photon pairs then propagate through a fiber of length $L$ with the same second-order dispersion parameter $\beta_2$ for both polarization modes, the JSA (Fourier transformed JTA) changes to $\beta (\omega_\text{i}, \omega_\text{s}) \propto \exp\left(-\frac{i}{2} \beta_2 L (\omega_\text{i}^2 + \omega_\text{s}^2) \right) \ F(\omega_\text{i}) F(\omega_\text{s}) \ J(\omega_\text{i}, \omega_\text{s})$ and remains separable in the sum/difference coordinates. It follows that the difference component of $\tilde{\beta}$ is given by the relatively simple formula
\begin{equation}
    \tilde{\beta}_-(\omega_-) = (\pi \rho)^{1/4} \exp\left( - \frac{\omega_-^2}{2 \rho} - \frac{i L \beta_2 \omega_-^2}{2}  \right),
\end{equation}
with $1/\rho = 1/r + 1/s$. Hence, evaluation of Eq.~\eqref{eq:coincidence_rate} for the given SPDC state leads to the coincidence rate
\begin{equation}
    \label{eq:calculated_coincidence_rate}
    \begin{split}
        &c(\tau) =\frac{1+\eta'}{4} \left[ \text{erf}\left(\sqrt{\frac{\rho'}{2}} (T+\tau) \right) + \text{erf}\left(\sqrt{\frac{\rho'}{2}} (T-\tau )\right) \right] \\ 
        - &\frac{1-\eta'}{2} \exp \left( -\frac{\rho}{2}  \tau ^2 \right) \text{Re} \left[ \text{erf}\left(\sqrt{\frac{\rho'}{2}} T+i \sqrt{\frac{\rho-\rho'}{2}} \tau \right) \right] \raisetag{5ex}
    \end{split}
\end{equation}
where $\rho' = \rho/ (1 + (L \beta_2 \rho)^2)$ signifies the broadened width after dispersion and $\eta' = (2\eta - 1)^2$ is a new beam splitter constant that is symmetric around $1/2$. 

\subsection{Implications}
A remarkable consequence of the time-resolved description is that sensitivity to identical dispersion for both photons is restored, as was also shown in \cite{Xia2023}. Usually, in the $T \rightarrow \infty$ limit, the delay imposed on one arm of the set-up commutes with the effect of dispersion -- i.e. $[\exp(-i \ L \ k(\hat{\omega}), \exp(-i\hat{\omega}\tau)] = 0$, where $k$ is some dispersion relation -- implying that the overlap of the two photons does not depend on dispersion. This is a manifestation of local dispersion cancellation and is contrary to the case of a finite coincidence window, which imposes an additional filter that does not commute with $\hat{\omega}$, hence generating dispersion-dependent behavior. In addition, our analytical prediction in Eq.~\eqref{eq:calculated_coincidence_rate} features oscillations as a function of $\tau$ due to the imaginary part of the error function. Their period can be found to be approximately $P_\text{osc.} = 2\pi/(T \sqrt{\rho' (\rho-\rho')})$. This result agrees with Steinberg's oscillations (in \cite{Steinberg1992} see equation 18 and identify $\rho \rightarrow 2\sigma^2$ and $\beta_2 \rightarrow P$).


\section{Methods}
\label{pa:exp_setup}
Photon pairs are generated via type-II SPDC in a periodically poled KTiOPO4 (ppKTP) crystal with a poling period of 46.2\,\text{\textmu m}, pumped by a pulsed 775\,nm Ti:sapphire laser. The crystal approximately satisfies the discussed EPM condition with group index differences $\Delta n_{g,s} = 0.0471$ and $\Delta n_{g,i} = -0.0415$, which are found by considering Sellmeier's equations with appropriate constants \cite{Fradkin1999, Emanueli2003, König2004}.

A 12\,nm FWHM bandpass filter is used after the ppKTP source to improve spectral purity. The orthogonally polarized signal and idler photons, centered at 1550\,nm, are separated using a polarizing beam splitter and coupled into polarization-maintaining (PM) fibers via fiber collimators mounted on piezoelectric translation stages for precise delay control.

The two photons were coupled into orthogonal polarization modes of the fiber under test (FUT) using a fiber polarization combiner. The FUT consists of single-mode SMF-28 spools of various lengths (1 - 5\,km), which can be concatenated to reach total fiber lengths up to 29\,km. Because SMF-28 is not polarization-maintaining, random birefringence induced polarization scrambling occurs during propagation. A fiber polarization controller at the output compensates this effect, restoring linear polarization states aligned with the input axes. Polarization alignment is optimized by maximizing single-photon count rates and minimizing cross-coupling between the arms.

At the output, the photons are separated using an identical fiber polarization combiner operated in reverse as a polarization splitter. The two output ports are connected to a PM 50:50 fiber beam splitter, ensuring that both photons enter with matched polarization states. The beam splitter outputs are directed to superconducting nanowire single-photon detectors. Detection signals are recorded by a time-tagging module, which identifies coincidence events based on programmable coincidence windows.

For each fiber length, coincidence counts are measured as a function of relative optical delay to obtain the HOM interference curve. Measurements are repeated for different fiber lengths and coincidence window widths to evaluate the effect of the coincidence window on the interference visibility and dip profile.

\begin{figure}[b]
    \centering
    \includegraphics[width=0.8\columnwidth]{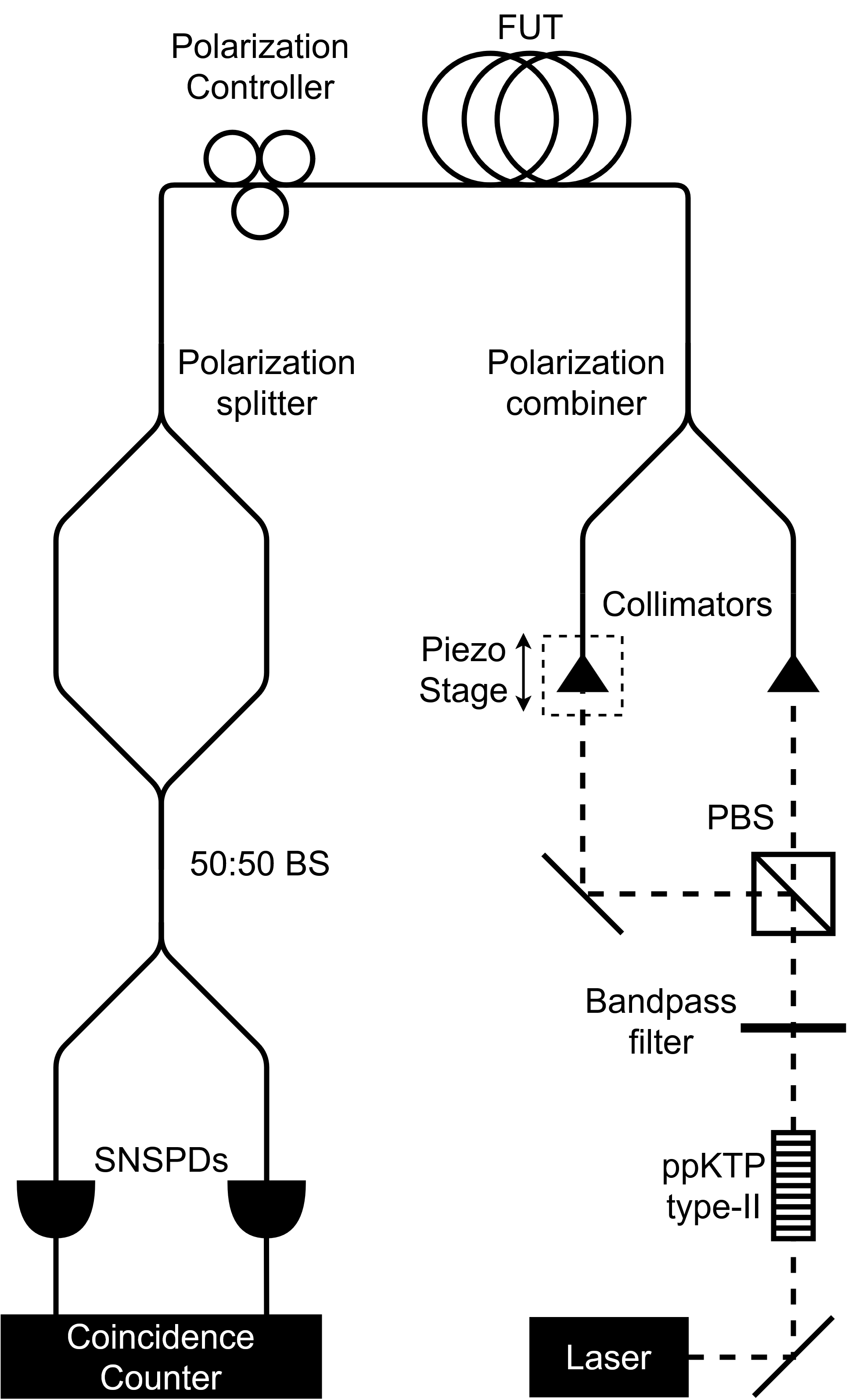}
    \caption{Orthogonally polarized photons from a filtered ppKTP source are separated and coupled into the FUT via a polarization combiner, with one arm delayed by a piezo stage. After propagating through the FUT, a polarization controller and splitter restore and separate the modes. The photons then interfere at a 50:50 beam splitter and are detected by SNSPDs connected to a coincidence counter.}
    \label{fig:setup}
\end{figure}


\section{Results}
To validate the theoretical model derived in Eq.~\eqref{eq:calculated_coincidence_rate}, we performed a global fit to the experimental dataset across 60 different configurations of coincidence window $T$ and fiber length $L$. The free parameters in the fit are the group velocity dispersion $\beta_2$, the spectral width parameter $\rho$, and the beam splitter reflectivity $\eta$, which is allowed to vary per dataset to account for alignment errors.

The obtained global fit yields a dispersion parameter of $\beta_2 = 21.39 \pm 0.02$\,ps$^2$\,km$^{-1}$ and a spectral parameter $\rho = 14.53 \pm 0.06$\,ps$^{-2}$. These values are consistent with the manufacturer's specifications for SMF-28 fiber ($\beta_2^\text{max} < 23$\,ps$^2$\,km$^{-1}$). Using the relation between the spectral parameters and the group index, we estimate the group index difference $|\Delta n_\text{g,a}|$ to range between $0.031$ and $0.064$, depending on the effective filter bandwidth assumed.

\begin{figure}[!tbh]
    \centering
    \subfloat[Fit for $T = 0.4$\,ns, $L = 10$\,km. \label{fig:fit1}]{%
        \includegraphics[width=0.49\columnwidth]{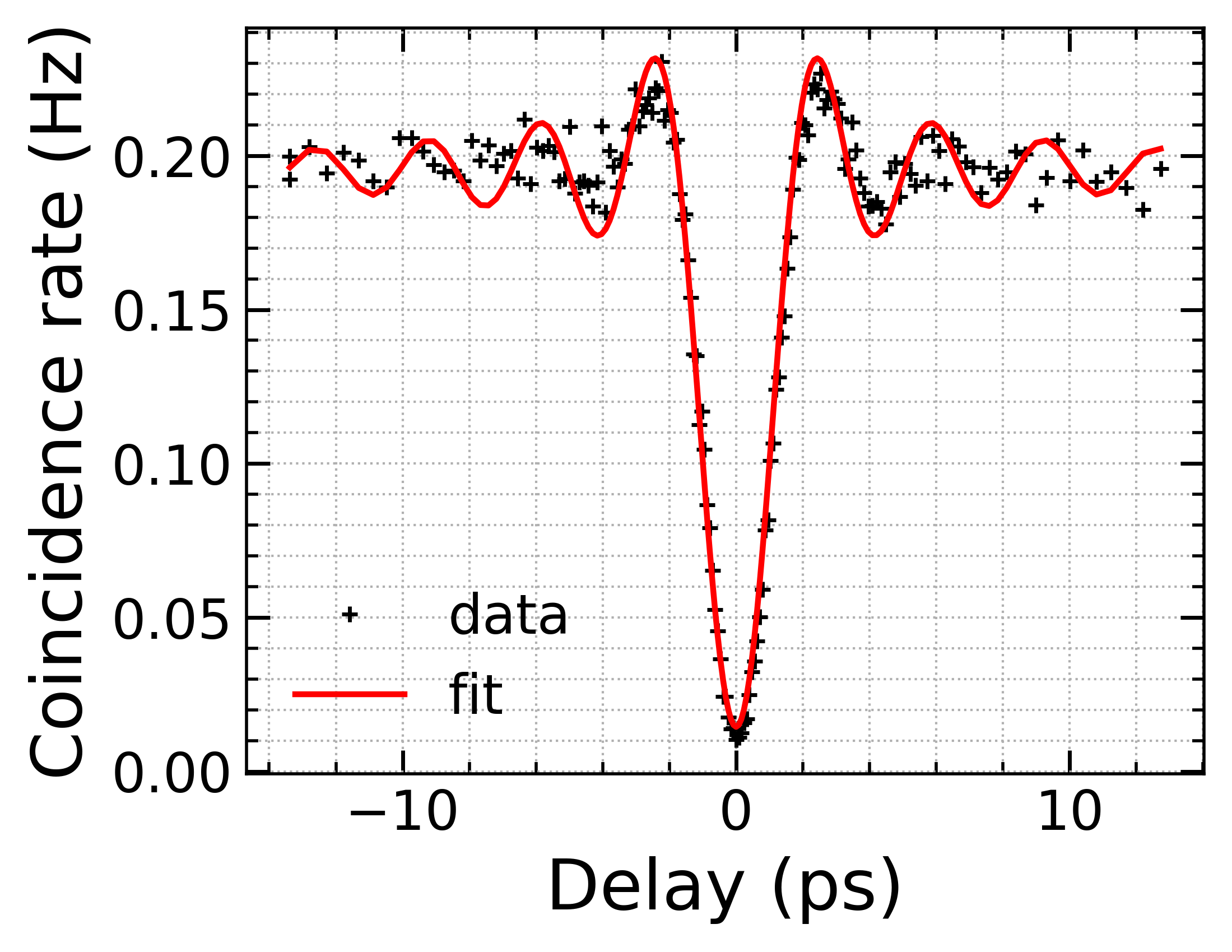}}%
    \hfill
    \subfloat[Fit for $T = 0.8$\,ns, $L = 10$\,km. \label{fig:fit2}]{%
        \includegraphics[width=0.49\columnwidth]{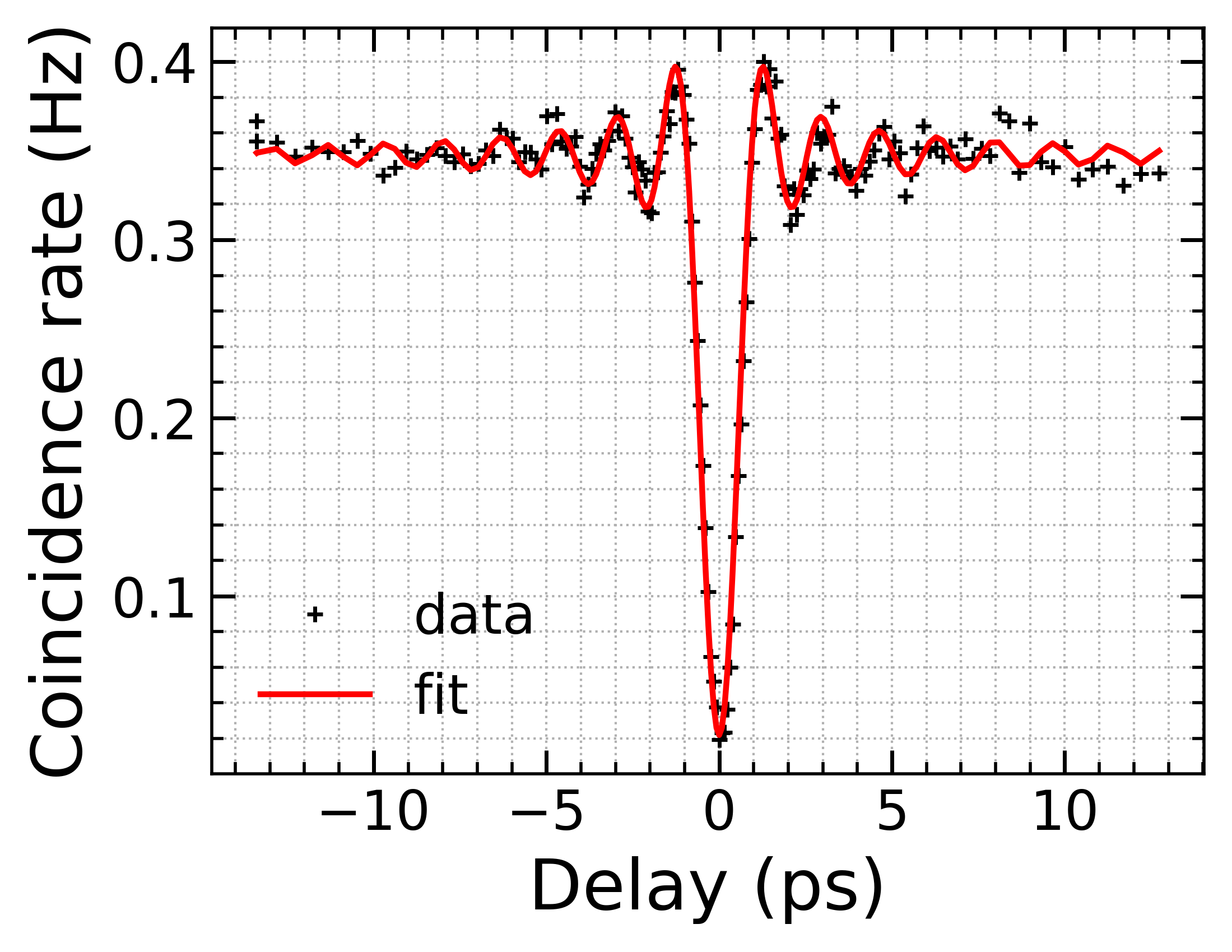}}
    
    \subfloat[Fit for $T = 1.0$\,ns, $L = 15$\,km. \label{fig:fit3}]{%
        \includegraphics[width=0.49\columnwidth]{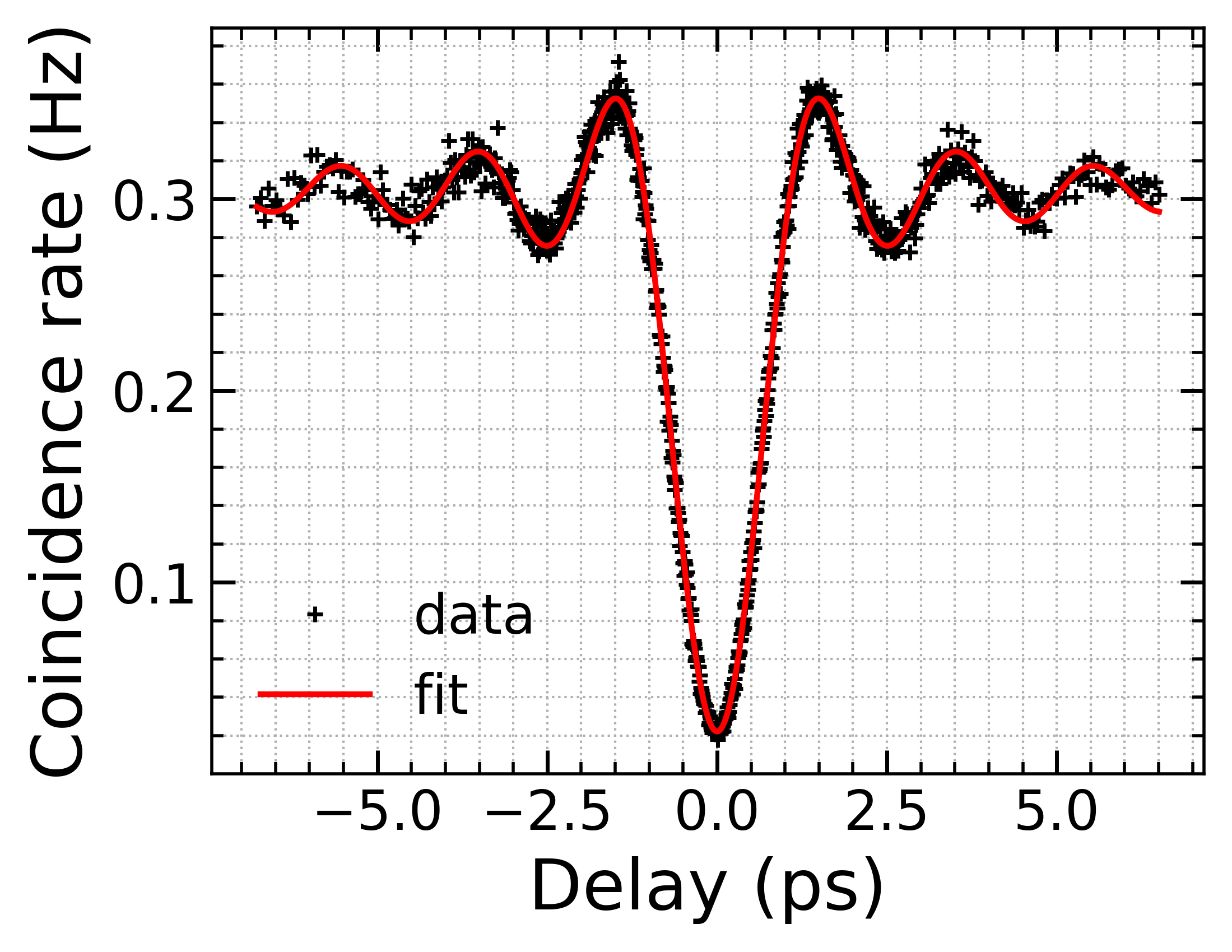}}%
    \hfill
    \subfloat[Fit for $T = 1.0$\,ns, $L = 20$\,km. \label{fig:fit4}]{%
        \includegraphics[width=0.49\columnwidth]{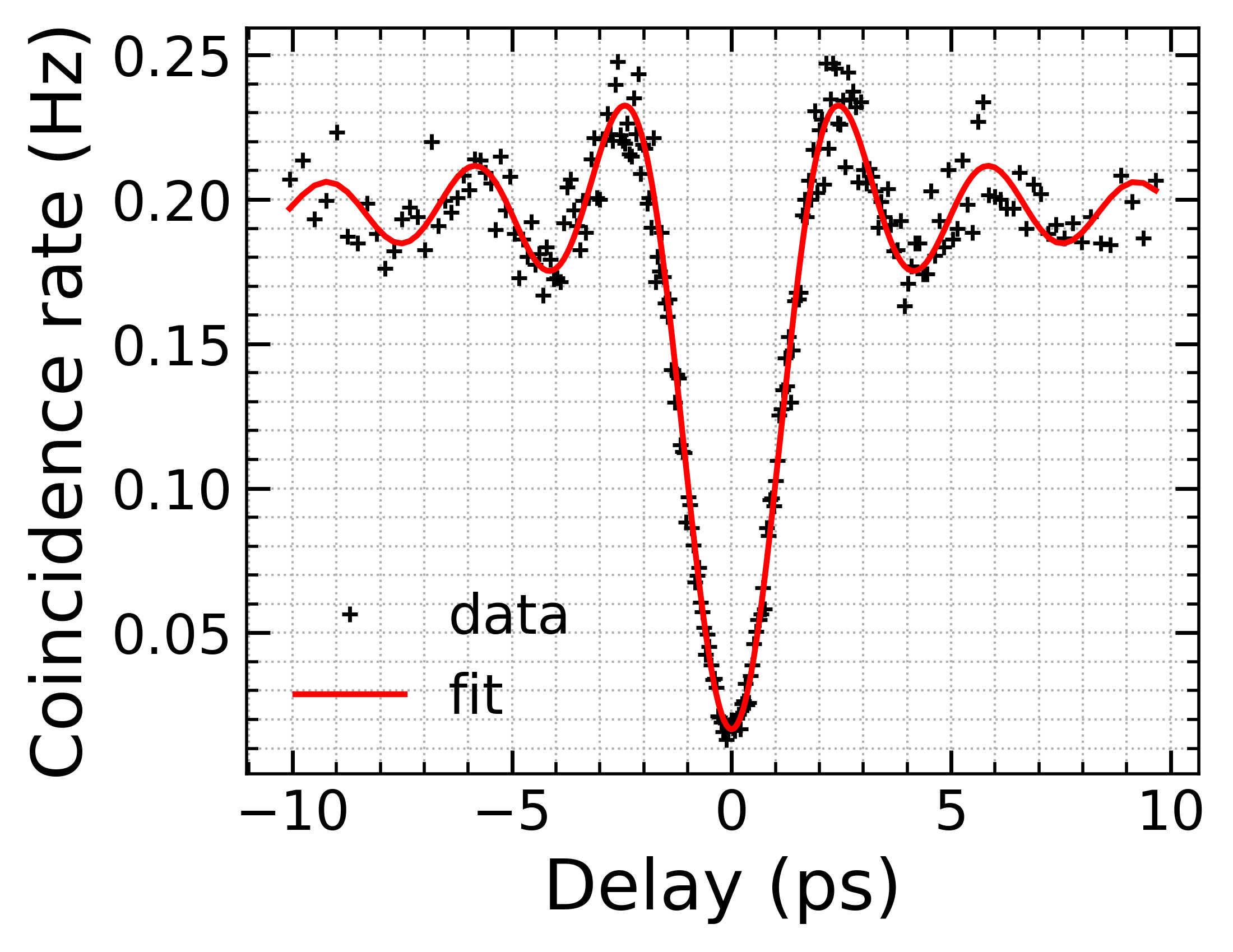}}
    
    \caption{Example fits showing predicted oscillations for different experimental configurations of coincidence windows $T$ and fiber lengths $L$.}
    \label{fig:example_fits}
\end{figure}

Figure~\ref{fig:example_fits} shows representative data for different fiber lengths and window lengths. The influence of the rectangular window is clearly visible: rather than a smooth Gaussian dip, the data exhibit characteristic side-lobes (oscillations). These oscillations become more pronounced as the ratio of the dispersive broadening to the window length increases, a signature of the sharp cut-off imposed by the time-tagger.

\begin{figure}[!tbh]
    \centering
    \includegraphics[width=\linewidth]{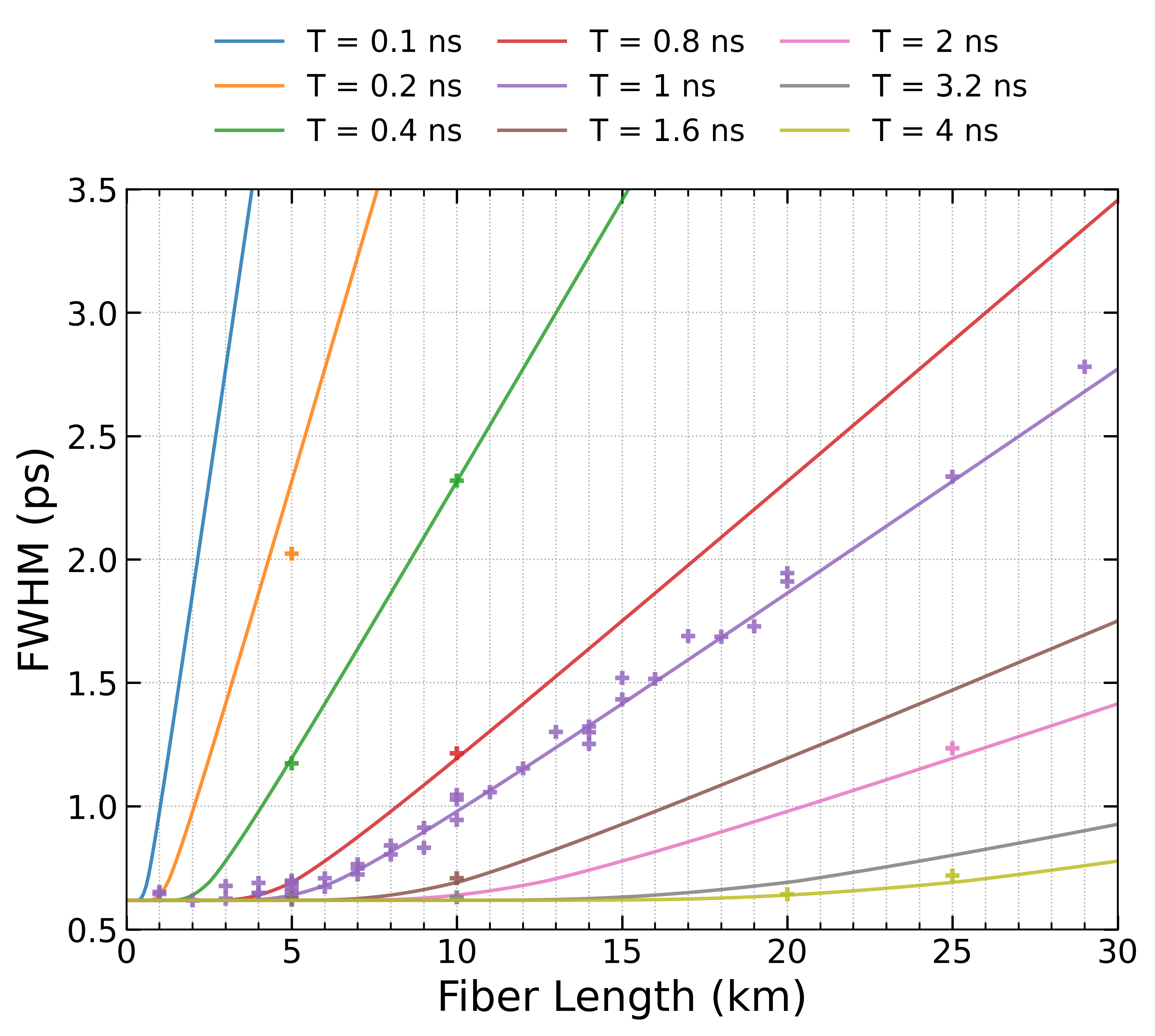}
    \caption{FWHM of the model (solid lines) and data (points) plotted against the fiber length for several values of the coincidence window. The model uses the extracted fit parameters, $\rho$ and $\beta_2$. While the fit allowed $\eta$ to vary, the model curves here are plotted with fixed $\eta=0.5$ to illustrate the theoretical trend.}
    \label{fig:fwhm}
\end{figure}

To further quantify the agreement, we extracted the full width at half maximum (FWHM) of the dips. As shown in Fig.~\ref{fig:fwhm}, the FWHM remains constant for short fiber lengths before transitioning to a linear regime. The onset of this transition and the subsequent slope are dependent on the coincidence window length $T$, marking the regime where the photon width exceeds the length of the coincidence window. The model (solid lines) accurately captures this broadening behavior apart from some deviations. Minor discrepancies for $T > 0.3$\,ns are attributed to slight variations in $\eta$ mentioned above. In the $T < 0.3$\,ns regime, a systematic error is observed (see Appendix \ref{ap:residuals}). This is likely due to the finite timing jitter of the SNSPDs ($\approx$ tens of ps), which becomes comparable to the window length $T$ in this limit, and effectively blurs the sharp rectangular window edges assumed in the model. Furthermore, while the model assumes an ideal system, other non-idealities, such as residual PMD obtained in the fiber, multi-pair emission from the SPDC source, and detector dark counts, may also contribute to minor deviations or visibility degradation.

\section{Conclusion \& Discussion}
We theoretically derived the influence of a rectangular coincidence window on Hong-Ou-Mandel interference. The extended phase-matching conditions of the ppKTP crystal allowed us to derive a simplified analytical expression for the coincidence rate presented in Eq.~\eqref{eq:calculated_coincidence_rate}. The model reveals that the finite coincidence window restores sensitivity to symmetric group velocity dispersion, an effect typically canceled in the infinite-window limit. This manifests as a reduction in signal intensity, dip broadening, and the emergence of characteristic oscillations, all of which were clearly resolved in our measurements over 29\,km of fiber. Notably, we derived the characteristic oscillatory period induced by the rectangular filter, finding agreement with Steinberg's predictions \cite{Steinberg1992}.

The experimental data shows excellent agreement with the phenomenological model, yielding a dispersion parameter of $\beta_2 = 21.39 \pm 0.02$\,ps$^2$\,km$^{-1}$, consistent with manufacturer specifications. The validation holds across the dispersive regime, with minor deviations appearing for $T<0.3$\,ns. This might be attributed to the timing jitter of the detectors as in this case the window length approaches the instrumental jitter. Moreover, the manufacturer group index difference, $\Delta n_{g,a}$, falls within the range found experimentally, suggesting a valid approach. The main influence on the experimental uncertainty for this quantity is the choice of effective filter bandwidth, $s$. The two bounds correspond to either matching the (originally rectangular) filter on respectively field or intensity level.

While the interplay between dispersion and temporal filtering has been explored before, previous models relying on Gaussian filters or infinite windows do not fully capture the specific hardware response of modern quantum communication networks. These systems increasingly rely on commercial time-tagging modules with rectangular coincidence windows, coupled with low-jitter detectors operating over highly dispersive, long-haul fiber links. In this regime, the sharp rectangular cut-offs emerge as a dominant, measurable effect. Our source-to-measurement model successfully captures this reality, providing a robust diagnostic framework for analyzing coincidence-window-induced dispersion effects. In conclusion, properly accounting for these hardware-specific dynamics is an essential step toward long-distance HOM-based quantum communication infrastructure.


\section*{Data Availability}
The data and code that support the findings of this study are openly available \cite{Hasenack2026}.


\begin{acknowledgments}
We thank G.N. Warnars for experimental support. We thank Chigo Okonkwo for lending fibers and components. Part of this work was financially supported by the QuantERA HSM-QCC project, the Challenges in CyberSecurity (CiCS) Zwaartekracht program and Groeifonds project Quantum Delta NL KAT-2.
\end{acknowledgments}

\hfill
\appendix

\section{Fitting procedure}
\label{ap:fitting_procedure}
Because the scale of the coincidence rate of the data depends on the laser intensity, whereas the calculated coincidence rate assumes exactly two photons, we need a means of comparison. The amplitude of the data points cannot just be normalized, since the maximum of the prediction depends on the fit parameters. Thus, we settle on a scale-agnostic error-metric
\begin{equation}
    \mathbbm{E} = \min_{s_i} \sum_x |s_i \ f(x) - y_x |^2 = \sum_x \left| s_{i,0} \ f(x) - y_x\right|^2
\end{equation}
where $s_i$, as $\eta_i$, again depends on the specific experimental realization and $s_{i,0} = (\sum_x f(x) y_x)/(\sum_x f(x)^2)$. Then, the data is fitted with the least-squares Levenberg-Marquardt algorithm, where error propagation is handled by the usual Jacobian estimation, $\text{cov} \approx  (J^\top J)^{-1} \ \sum_x r_x^2/(n-p)$, where $r_x$ are the scale-agnostic residuals and $p$ is the number of fit parameters. The manual way of calculating the uncertainty in this way ensures that the dependence of $s_{i,0}$ on the fit parameters is accounted for.

\section{Residuals}
\label{ap:residuals}
\begin{figure}[H]
    \centering
    \includegraphics[width=\columnwidth]{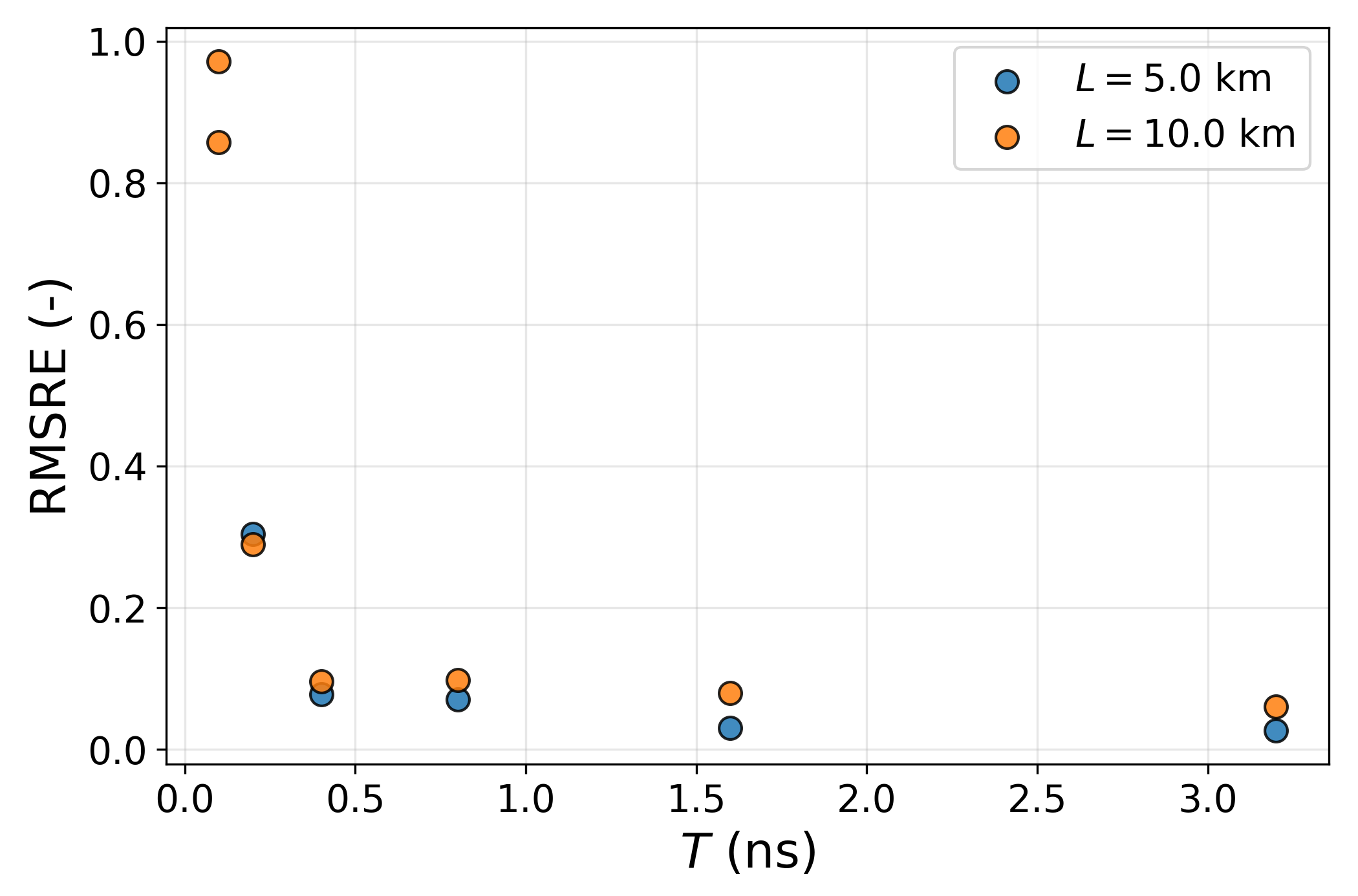}
    \caption{Average scaled residuals of a subset of the data with respect to fiber length and coincidence window.}
    \label{fig:residuals}
\end{figure}

To assess the validity of our theoretical model across different experimental regimes, we analyze the quality of the fits by examining the residuals. Figure~\ref{fig:residuals} displays the average root mean square relative error, $\text{RMSRE} = \sqrt{\sum_x (r_x/y_x)^2/N}$, for a subset of the data. The overall small scaled residuals give confidence in the model. For $T <0.3$\,ns, the RMS error exhibits an upward trend above the noise floor.


\clearpage
\bibliography{references}

\end{document}